# Recent Development in Disease Diagnosis by Information, Communication and Technology

SHABANA UROOJ[1], ASTHA SHARMA[2], CHITRANSH SINHA[3], FADWA ALROWAIS[4]

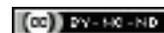

## ABSTRACT

The usage of Information, Communication and Technology (ICT) in health sector has a great potential in improving the health of individuals and communities, disease detection, prevention and overall strengthening the healthcare systems, vital for development and poverty reduction. Large ICT establishments offer a variety of Artificial Intelligence (AI) based solutions; and their tenacities are inclusive of wearable therapeutic devices, healthcare management arrangements, extrapolative healthcare diagnostics, ailment prevention systems, detection and screening of diseases and automated tactics. In the field of healthcare related instrumentation, AI plays a prevalent role with the amalgamation of several technological progressions. This enables machines to sense, comprehend, act and learn to perform organisational and clinical healthcare functions as well as serves the research and training purposes. Additionally, it enables to accomplish the anticipated directorial and medicinal benefits. The major causes of life threats reported in literature are; heart and brain diseases. In this paper, an extensive review is presented exploring the evolving ICT technologies in machine learning and AI to help ICT enthusiasts to be able to catch up with the emerging trends in healthcare.

## INTRODUCTION

Recently, a booming interest is comprehended in wearable and implanted health monitoring technologies leading to transforming the healthcare systems. This has led to extensive adoption of Wireless Body Area Networks (WBANs) which facilitates ubiquitous patient monitoring by allowing self-management of diseases through partnerships between patients and healthcare professionals. This aids in overall proactive wellness management by providing early detection and prevention of diseases [1]. WBAN exploits wireless communication technology to gather real-time data from a network of independent, intelligent, low-power, miniaturised sensors and actuators positioned on the surface of body or invasive implants in the body. The standards popularly adopted for medical WBAN are IEEE 802.15.6 and IEEE 802.15.4j [2,3]. However, new protocols are required in WBANs designing in comparison to the utilised in generic goal Wireless Sensor Networks (WSN).

Information and Communication technologies have enabled disease diagnosis by monitoring patient symptoms and sharing the information globally with scientific community for effective management of data. Generally, there are mainly two different approaches to designing disease diagnosis algorithms in various domains. One being an Expert System (ES) or Knowledge Based System (KBS) which encodes the knowledge of experts and clinical knowledge gathered based on evidence into a form of algorithm. It comprises of four majors: (i) a knowledge base consisting of diseases, findings and relationships; (ii) an inference engine; (iii) knowledge engineering tool and (iv) a specific user interface which is shown in [Table/Fig-1] [4,5]. The salient features of classical ES are majorly interactive training tool, decision support system and expert advice.

However, it suffers from certain issues such as lack of extensibility, time-consuming process of creation of knowledge base, difficulty in structuring an algorithm, need of collaboration of experts etc., [6,7]. This has led to a trending shift from rule-based approach to a data-driven approach paving the way for a new era of data analytics and AI. Experts systems can benefit from data mining techniques to form knowledge base.

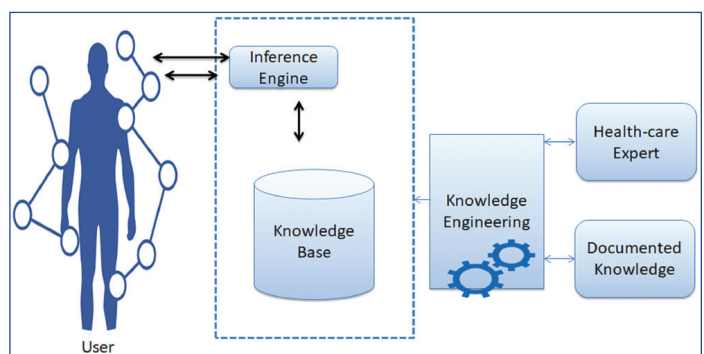

**[Table/Fig-1]:** A classical expert system for disease diagnosis.

The other approach employs data mining tools which transform the medical data into new knowledge where machine learning algorithms are applied to large databases [8]. Although data mining techniques have been applied to various domains, its application in the field of healthcare is seen to be evolving with the growing interest of academicians, researchers and medical practitioners. This is because the healthcare domain faces the challenge of complexity due to numerous medical data standards, increasing amounts of unstructured data, patient privacy constraints and ambiguous semantics. Further, the facility of reduced storage cost and connectivity have made it possible to have very large medical datasets available over the Internet which has allowed to run various complex machine learning algorithms by exploiting cheaper computation [9-11].

The statistical theories are used by data mining algorithms to build mathematical models in order to make inference from a sample. The goal of building algorithms is to optimise the parameters of the model using the training data or past experience in order to help the machine learn. This model can be predictive/supervised or descriptive/unsupervised based on whether the application is to make predictions in the future or to gain knowledge from data accordingly. In supervised learning approach, the aim is to learn a mapping from the input to an output from a set of training input-output pairs. On the other hand, only input is provided in unsupervised learning where the aim is to find interesting patterns in the data for knowledge discovery. In case of disease diagnosis, the relevant information about the patient gathered with help of WBAN sensors is the inputs and the different diseases diagnosed are the classes. A typical disease diagnosis system employing machine learning algorithms for classification or prediction of disease using feature vectors of disease and symptoms is shown in [Table/Fig-2].

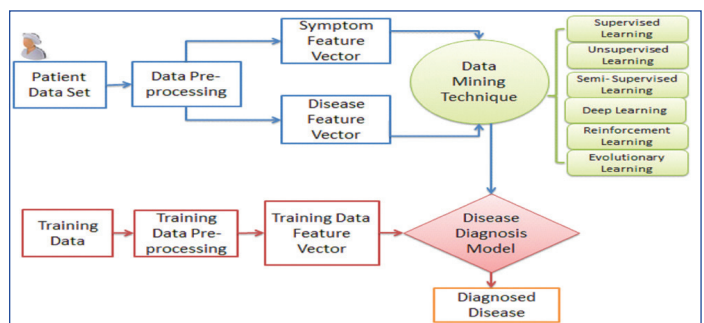

**[Table/Fig-2]:** A disease diagnosis system employing machine learning algorithms for prediction.

It is well known that supervised learning use only characterised data while unsupervised learning utilises unlabelled data. Another popular learning technique is semi-supervised which have the ability to improve the learning accuracy by using fewer labelled data points in combination with huge amount of unlabelled data points. Since innovation in the machinery and technologies has always been inspired by human brain; it has led to emergence of a subfield of machine learning termed as Deep Learning (DL). DL practices Artificial Neural Networks (ANNs) to automatically extract the pertinent features from a huge set of labelled data and many numbers of abstraction hidden layers instead of manually choosing the feature vectors as is done in traditional machine learning. Reinforcement learning involves finding the best possible behaviour or path in case of specific situation by exploring various possibilities and can perform better in more ambiguous real-life environments. Evolutionary learning is inspired by biological evolution and involves generation of a set of possible candidate solutions where the less desired solution is dropped with each iteration or generation [12].

The [Table/Fig-3] shows the top 10 leading causes of death worldwide. Amongst them, the major percentage responsible for death reported is mainly due to heart [13] and brain diseases [14-16]. This serves as motivation for the paper to present a comprehensive literature review on recent trends witnessed in the field of data mining for the heart and brain diseases respectively. The present authors have utilised the research trend observed in the year from 2016 to 2018 for the present work. In the case of brain diseases, numerous works existed where advancement in machine learning algorithms such as genetic algorithms, fuzzy systems, neural networks, deep learning etc., have been already explored for brain

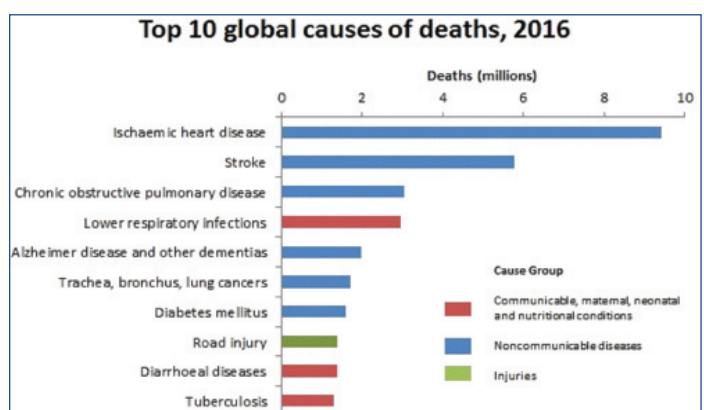

**[Table/Fig-3]:** Top 10 leading causes of death according to Global Health Estimates (GHE 2016).

diseases prediction. Therefore, the present authors have basically focused on the key trends observed in the year 2018 which will help readers to understand and grow their interest in the evolving ICT technologies in the field of medical diagnosis.

## LITERATURE REVIEW

### Ischaemic Heart Diseases

Cardiovascular diseases are the leading cause of death worldwide for the past 10 years accounting for nearly 17.9 million deaths (44% of all non-communicable diseases deaths) [17]. Amongst all the cardiovascular diseases, ischaemic or coronary heart disease is the most common which occurs when the cholesterol particles cause blockage of blood flow due to its accumulation on the arteries walls. This decrease in flow of blood causes reduction in amount of oxygen supplied to heart muscle.

Fortunately, ischaemic heart disease can be prevented and controlled by adopting lifestyle changes, medication and surgical procedures. If left unchecked, it may lead to severe heart damage and may even lead to life threatening heart attack. The most common symptoms include chest pain, indigestion, heartburn, weakness, nausea, cramping and shortness of breath.

The role of ICT can be witnessed in the development of an accurate and systematic tool for identifying the high-risk patients and gathering data for timely diagnosis [18,19]. Numerous literature works exist on diagnosis of heart diseases and adopting various data mining techniques for its prediction [20,21]. The most common dataset used for analysis has been Cleveland heart disease dataset. Most of the existing work and some of the recent work has focused on determining the performance of individual data mining techniques and providing an overall comparison between them [22-24]. The measures adopted for quantifying the performance are Accuracy, Sensitivity and Specificity.

(1) $$Accuracy = \frac{(TP+TN)}{TP+TN+FP+FN}$$

(2) $$Sensitivity = \frac{TP}{TP+FN}$$

(3) $$Specificity = \frac{TN}{TN+FP}$$

where TP is true positive, TN is true negative, FP is false positive and FN is false negative. From the [Table/Fig-4], it is evident that Support Vector Machine (SVM) emerged as a popular choice among the several adopted data mining techniques [22-25,26,27]. [Table/Fig-5] depicts the trend observed in various techniques adopted along with their merits and demerits [23,24,27-28].

| References | Year | Data mining technique adopted | Accuracy (%) | Sensitivity (%) | Specificity (%) |
|---|---|---|---|---|---|
| Nashif S et al., [22] | 2018 | SVM | 97.53 | 97.50 | 94.94 |
| | | Random forest | 95.76 | 95.80 | 92.61 |
| | | Simple logistic | 95.05 | 95.10 | 93.77 |
| | | Naïve bayes | 86.40 | 86.40 | 82.88 |
| | | ANN | 77.39 | 77.40 | 63.42 |
| Malav A and Kadam K, [23] | 2018 | Hybrid of ANN and K-means | 93.52 | 89.83 | 90.98 |
| Assari R et al., [24] | 2017 | SVM | 84.33 | 78.26 | 89.5 |
| | | Naïve bayes | 83.66 | 80.43 | 86.41 |
| | | KNN (K=7) | 81 | 77.53 | 83.95 |
| | | Decision tree | 79 | 76.08 | 81.48 |
| Khan SN t al., [25] | 2017 | SVM | 84.12 | 90.00 | 77.20 |
| | | RIPPER | 81.08 | 86.25 | 75.82 |
| | | ANN (MLP) | 80.06 | 83.75 | 75.73 |
| | | Decision tree C4.5 | 79.05 | 83.12 | 74.26 |
| Gupta N et al., [26] | 2017 | Naïve bayes | 86.42 | 83.7 | 89.5 |
| | | AdaBoost | 86.21 | 85.7 | 86.4 |
| | | Boosted tree | 85.75 | 83.1 | 84.9 |
| | | SVM | 85.18 | 81.4 | 89.5 |
| | | Binary discriminant | 84.26 | 97.2 | 96.3 |
| | | MLP | 83.95 | 83.7 | 84.2 |
| | | Random forest | 81.48 | 74.4 | 89.5 |
| | | J48 | 77.78 | 62.8 | 94.7 |
| Bala A et al., [27] | 2016 | Naïve bayes | 86.42 | | |
| | | MLP | 83.95 | | |
| | | Random forest | 81.48 | | |
| | | J48 | 77.78 | | |

**[Table/Fig-4]:** Latest trends of various data mining techniques adopted for disease diagnosing with their performance measures [22-25,26,27].

| Technique | Description | Advantages | Disadvantages |
|---|---|---|---|
| SVM | SVM classifies the data into two major classes by finding an optimal hyperplane which is constructed by maximising the margins between these two classes. It is a linear as well as a non-linear classifier. | Linear SVM segregates the data points by a linear decision boundary. Has good generalisation capability. | For complex dataset where it becomes difficult to classify by linear SVM, in such case the data is mapped into high-dimensional feature space using kernels. Afterwards, it determines the optimal classification hyperplane in high-dimensional feature space. It suffers from the main problem of selection of suitable kernel function parameters, high algorithmic complexity and slow speed in testing phase |
| Hybridisation of data mining techniques [23] | Here, the pre-processed data is first clustered using k-means algorithm which is then given as input to the ANN classifier. | Reduced cost, increased accuracy and improved efficiency of prediction. Faster convergence by k-means. | Longer training times by ANN classifier. |
| Ensemble learning method [24] | It is employed to compare the performance of using combination of different data mining techniques. Generally, ensemble method includes some machine learning algorithms to produce one optimal predictive model. It constructs a set of hypotheses and combines them to form a single hypothesis in order to learn from training data. | It is usually employed to reduce variance (Bagging), decrease bias (Boosting) or to modify predictions (Stacking). Literature shows that the highest accuracy of 84.15% is achieved by using a combination of SVM and Multilayer Perceptron (MLP) techniques. Ensemble learning has stronger generalisation ability in comparison to a single learner. | However, there are certain issues such as lack of comprehensibility and lack of diversity measure which it still needs to explore. |
| Cloud computing technology in healthcare domain [28] | Cloud computing is a subscription-based service where different computing services (storage, servers, databases, software, networks, analytics etc.,) are delivered on the payment basis over the Internet or the 'cloud'. | Its high demand is due to the soaring usage of healthcare Internet-of-things (IoT) devices and the necessity of increased memory and computational requirements. It enables the usage of key application technologies such as big data analytics, cognitive computing, mobile collaboration and information sharing. | |
| | In [26], an intelligent decision support model is developed on cloud environment to improve the heart diseases prediction based on present vital patient statistics. To enhance the accuracy of prediction and reduce execution time, the top three models (Naïve Bayes, Adaboost and Boosted tree) are ensembled on cloud platforms and their performances are compared with that of standalone machine. | Results show significant gain achieved by using cloud platforms over the standalone machine where different performance measures such as kappa statistics, standard deviation, execution time, classification accuracy etc. were employed to validate the results. | |
| Green ICT-based prognosis model based on cloud computing technology [27] | The authors have determined the performance of various individual data mining technique on standalone machine environment and then used that technique which gave highest accuracy to predict on cloud environment. | From observations, it is found that Naïve Bayes yields highest accuracy of 86.42% in standalone machine environment and an improved accuracy of 88.89% in cloud environment, thus advocating the usage of green ICT. | |

**[Table/Fig-5]:** Recent trends observed in AI techniques adopted for diagnosing heart diseases along with their respective merits and demerits [23,24, 27-28].

## Brain Diseases

Human brain, a massively complex information processing system can provide an insight into developing new treatments for brain diseases, build robust artificial intelligence techniques and improves the overall socio-economic life of people. Some of challenging ailments dealing with vast array of unknown neurons, various brain disorders diseases such as Alzheimer's, Parkinson's, Autism Spectrum Disorder, Schizophrenia, multiple sclerosis and many more [29-31] needs serious attention.

Large amount of genetic, phenotypic, imageing and behavioural data are gathered for numerous brain diseases in order to facilitate the invention of new biomarkers for effective disease diagnosis. However, detection of brain diseases presents certain challenges such as data gathered being high-dimensional, noisy, unstructured, have non-linear separability and heterogeneity and lack of symptoms detection in early stages or similarity in detected symptoms among multiple brain diseases. Advancement in ICT technologies such as neuroimaging, medical informatics and neurocomputing can play a key role in systematically investigating brain diseases, capturing early cognitive changes and achieving individual level disease diagnosis [29,30].

The [Table/Fig-6] presents the recent trends of various data mining methods adopted for prediction or classifying the distinct types of brain ailments [24,25,26,32-34]. It can be observed that deep learning neural network has attained widespread attention of researchers as it is able to automatically learn increasingly abstract features through a multi-layered hierarchical structure. Further, it is able to learn understated hidden patterns from high-dimensional neuroimaging data and efficiently capture the disease-related pathologies.

A data-driven classification using SVM is applied to Thalamocortical Dysrhythmia (TCD) model to explain divergent neurological disorders and to investigate spectral equivalence between different neurological (tinnitus, pain and Parkinson's disease) and neuropsychiatric (depression) by authors Bala A et al., [27]. Results show the significance of theta, beta and gamma frequency bands in differentiating between neuropsychiatric disorders and healthy control subjects. Spatially distinct brain areas were selected to discriminate between the different clinical TCD entities. The study supports the validity of TCD as an oscillatory mechanism present in divergent neurological disorders. A detailed description of the techniques cited in [Table/Fig-6] is presented along with their performance results in [Table/Fig-7] [22,23,25,26,27,32].

| Ref. No. | Technique adopted | Description |
|---|---|---|
| Assari R et al., [24] | Deep Learning Algorithm | The authors have applied DL algorithms to detect ASD among vast brain imaging dataset taken from a world-wide multi-site database called ABIDE (Autism Brain Imaging Data Exchange) determined from patient's brain activation patterns. Although usage of huge multi-site datasets is challenging due to variations introduced by different subjects, scanning measures and apparatus which results in addition of noise in datasets leading to difficulty in classification. |
| Cai H et al., [32] | Complex Network model+Random Forests classifiers+SVM | A complex model is built to identify those regions which are severely affected by Parkinson's disease and then associate them with an easy to interpret statistical significance level as connectivity measures. Random forest classifiers are employed as a package for feature selection and to learn a compact representation. After which, SVM is used to link complex network features with clinical scores to provide a diagnostic index. The results show enhanced discriminative power of clinical features using complex model. |
| Khan SN t al., [25] | Random Neural Network (RNN) cluster | RNN exploits the features of both NNs and probabilistic queueing theory-based models and requires the usage of optimisation algorithms for training. Amongst the chosen five types of NNs, Elman NN is selected as optimal base classifier for feature selection and to find the abnormal regions. |
| Bruna J et al., [33] | Graph Neural Networks (GNNs) | A Graph Convolutional Network (GCN) is a GNN that enables CNNs to directly operate on graphs and updates feature vectors of nodes based on the properties of their neighbourhoods. GCN was first introduced by Bruna J et. al. [33] while the first application of GCN for brain examination in populaces and its diagnosis has been proposed by Parisot S. et. al [34]. A GCN based generic framework that controls imaging and non-imaging data to evaluate the outcome of distinct element on prediction of disease [26]. Different graph structures and baselines are explored showing the benefit of utilising phenotypic graph formulations in getting more accurate and stable results and the significance of choosing suitable phenotypic measures required for modelling pairwise interactions. |

**[Table/Fig-6]:** Recent trends observed in AI techniques adopted for diagnosing brain diseases [24,25,26,32-34].

| Reference | Type of disease | Technique adopted | Data Set | Outcome | Interpretation | Significance |
|---|---|---|---|---|---|---|
| Nashif S et al., [22] | Schizophrenia | Deep learning based – cross- site transfer classification Developed a deep Discriminant Autoencoder Network with Sparsity constraint (DANS) for automatic diagnosis of schizophrenic patients. | A large multi-site functional MRI sample (n=734 including 357 schizophrenic patients from 7 imaging resources) | Accuracy of approx. 85% (in multi-site pooling classification) and 81% (in leave-site-out transfer classification) | Findings show that dysfunctional integration of the cortical-striatal-cerebellar circuit across the default, ventral attention/salience and frontoparietal control networks play an important role in dysconnectivity model. | The proposed discriminant deep learning method may be able to learn reliable connectome patterns and aid in achieving accurate prediction of disease across multiple independent imaging sites. |
| Malav A and Kadam K, [23] | Autism Spectrum Disorder (ASD) | Deep learning algorithms | ASD patients brain imaging data from ABIDE (Autism Brain Imaging Data Exchange) database | Accuracy of 70%, Sensitivity of 74% and Specificity of 63% | The obtained patterns show an anticorrelation of brain function between anterior and posterior areas of the brain | Those areas of brain that contributed most to differentiating ASD from typically developing controls are identified using deep learning. Anticorrelation shows the empirical evidence of anterior-posterior disruption in brain connectivity in ASD. |
| Cai H et al., [32] | Parkinson's Disease (PD) | Used Complex Network model + Random Forests + SVM | Parkinson's Progression Markers Initiative (PPMI) database (169 normal controls and 374 PD patients) | Accuracy of 93%, Sensitivity of 93% and Specificity of 92% achieved using SVM classifier | Proposed methodology effectively detects which brain regions are mostly affected by the disease and uses network measures to provide a classification score. An unsupervised general methodology is used to model brain connectivity. | Showed that the proposed approach can learn an accurate model to discriminate control and patients, and is even able to detect possible novel imaging markers of the disease. Showed that use of MRI features becomes strategic for development of early diagnosis tools. |
| Khan SN t al., [25] | Alzheimer's disease | Random Neural Network | ADNI dataset | Accuracy of Elman neural network (NN) cluster could reach to 92.31% | Found 23 abnormal regions such as precentral gyrus, the frontal gyrus and supplementary motor area using Elman NN. | Random Elman neural network cluster could be used for selecting significant features which is, in turn able to determine abnormal regions. |
| Gupta N et al., [26] | ASD and Alzheimer's disease | Graph Convolutional Networks | ABIDE and ADNI (Alzheimer's Disease Neuroimaging Initiative) database | Accuracy of 70.4% (for ABIDE) and 80% (for ADNI) | Compared to other graph structures and baselines, phenotypic graph construction strategy yields more accurate and stable results as well as showed the importance of choosing appropriate phenotypic measures to model pairwise interactions. | Explored several graph structures and demonstrated the significance of using graph on classification accuracy. |
| Bala A et al., [27] | Thalamocortical dysrhythmia (TSD) | SVM learning | 541 subjects with 245 women and 296 men aged between 20-75 years; out of them 264 healthy control subjects, 153 tinnitus subjects, 78 subjects with chronic pain, 31 subjects with Parkinson's disease and 15 subjects with major depression. | Accuracy of 87.71% (tinnitus), 92.53% (pain), 94.34% (Parkinson's disease), 75.40% (depression) 87.60% (for full model including all four diseases) achieved using SVM in comparison to random model. | Results show that theta, beta and gamma-frequency bands are indispensable in distinguishing between neuropsychiatric disorders and healthy control subjects. Also, showed that there is spectral equivalence between neurological (i.e., tinnitus, pain, Parkinson's disease) and neuropsychiatric (i.e., depression) disorders with spatially distinct forms of TCD. | TCD model aids in explaining divergent neurological disorders. Affected brain areas were able to be identified. The study confirms the validity of TCD as an oscillatory mechanism underlying diverse neurological disorders. |

**[Table/Fig-7]:** Recent trends observed in various data mining techniques adopted for predicting or classifying the different types of brain diseases.

## CONCLUSION(S)

ICT will emerge as a leading technology that has the potential to revolutionise healthcare through remote monitoring, disease management and early disease detection. The other major issues to be taken into consideration for healthcare industry in future are; escalating costs, staff shortages, tightened regulatory requirements, increased burden of diseases, growing expectations of patients and improved efficiency. In future, it is required to be driven by mobility, big data analytics and IoT based devices that will need omnipresent services all the time at the highest efficiency.

This paper has provided a comprehensive review of growing technologies and research taking place in the field of diagnosis of heart and brain diseases. This review paper will help those readers who are new to the field and aspire to explore the role of ICT in disease diagnosis. This paper brings together the fragmented developments taking place in machine learning algorithms and artificial intelligence to application in the field of disease diagnosis, specifically in the heart and brain diseases.

## Acknowledgement

This research was funded by the Deanship of Scientific Research at Princess Nourah bint Abdulrahman University through the Fast-Track Research Funding Program.

## REFERENCES

[1] Lindberg B, Nilsson C, Zotterman D, Söderberg S, Skär L. Using information and communication technology in home care for communication between patients, family members, and healthcare professionals: A systematic review, Int J. of Telemedicine and Applications. 2013;461829. doi: 10.1155/2013/461829.
[2] Movassaghi S, Abolhasan M, Lipman J, Smith D, Jamalipour A. Wireless Body Area Networks: A survey. IEEE Communications Surveys & Tutorials. 2014;16(3):1658-86.
[3] Mehfuz S, Urooj S, Sinha S. (2015). Wireless body area networks: A review with intelligent sensor network-based emerging technology. Information Systems Design and Intelligent Applications. Springer. 2015:813-21.
[4] Biswas D, Bairagi S, Panse N, & Shinde N. Disease diagnosis system. Int J. of Computer Science & Informatics. 2011;1(2);48-51.
[5] Soltan RA, Rashad MZ, El-Desouky B. Diagnosis of some diseases in medicine via computerized experts' system. Int J. of Computer Science & Information Technology. 2013;5(5);79-90.
[6] Ravuri M, Kannan A, Tso G J, Amatriain X. Learning from the experts: From expert systems to machine learned diagnosis models. 2018.arxiv.org/abs/1804.08033.
[7] Chui K, Alhalabi W, Pang S, Pablos P, Liu R, Zhao M. Disease diagnosis in smart healthcare: Innovation, technologies and applications. Sustainability. 2017;9(12);2309.
[8] Bishop CM. Pattern recognition and machine learning (information science and statistics) springer-verlag new york. Inc. Secaucus, NJ, USA. 2006.
[9] Yoo I, Alafaireet P, Marinov M, Pena-Hernandez K, Gopidi R, Chang JF, Hua L. Data mining in healthcare and biomedicine: A survey of the literature. J. of Medical Systems. 2012;36(4);2431-48.
[10] Banaee H, Ahmed M, Loutfi A. Data mining for wearable sensors in health monitoring systems: A review of recent trends and challenges. Sensors. 2013;13(12);17472-500.
[11] Jothi N, Husain W. Data mining in healthcare-A review. Procedia Computer Science. 2015;72;306-13.
[12] Robert C. Machine learning, A probabilistic perspective. J. Chance, Taylor & Francis. 2014;27:2:62-63. doi: 10.1080/09332480.2014.914768 2014.
[13] Benjamin EJ, Virani SS, Callaway CW, Chamberlain AM, Chang AR, Cheng S, et al. Forecasting the future of cardiovascular disease in the United States: A policy statement from the American Heart Association. Circulation. 2018;137(12):e67-492.
[14] GBD 2016 Dementia Collaborators. Global, regional, and national burden of Alzheimer's disease and other dementias, 1990-2016: A systematic analysis for the Global Burden of Disease Study 2016. The Lancet Neurology. 26 Nov 2018. doi:10.1016/S1474-4422(18)30403-4.
[15] Global Health Estimates 2016: Deaths by Cause, Age, Sex, by Country and by Region, 2000-2016. Geneva, World Health Organization; 2018.
[16] World Health Statistics 2018: Monitoring health for the SDGs, sustainable development goals. Geneva: World Health Organization; 2018.
[17] https://www.who.int/news-room/fact-sheets/detail/the-top-10-causes-of-death.
[18] Purusothaman G, Krishnakumari P. A survey of data mining techniques on risk prediction: Heart disease. Ind J. of Science and Technology. 2015;8(12).
[19] Brahmi B, Shirvani MH. Prediction and diagnosis of heart disease by data mining techniques. J. of Multidisciplinary Engineering Science and Technology. 2015;2:164-68.
[20] Bhatla N, Jyoti K. An analysis of heart disease prediction using different data mining techniques. International Journal of Engineering. 2012;1(8);01-04.
[21] Patil SB, Kumaraswamy YS. Intelligent and effective heart attack prediction system using data mining and artificial neural network. European J. of Scientific Research. 2009;31(4);642-56.
[22] Nashif S, Raihan R, Islam R and Imam MH. Heart disease detection by using machine learning algorithms and a real-time cardiovascular health monitoring system. World J. of Engineering and Technology, 2018;6;854-73. ISSN Online: 2331-4249.
[23] Malav A, Kadam K. A hybrid approach for heart disease prediction using artificial neural network and K-means. Int J. of Pure and Applied Mathematics. 2018;118(8);103-10.
[24] Assari R, Azimi P, Taghva MR. Heart disease diagnosis using data mining techniques. Int J. Econ Manag Sci. 2017;6:415. doi: 10.4172/2162-6359.1000415.
[25] Khan SN, Nawi NM, Shahzad A, Ullah, A, Mushtaq MF, Mir J, et al. Comparative analysis of heart disease prediction. JOIV: Int J. of Informatics Visualization. 2017;1(4-2);227-31.
[26] Gupta N, Ahuja N, Malhotra S, Bala A, Kaur G. Intelligent heart disease prediction in cloud environment through ensembling. Expert Systems, 2017;34(3):e12207.
[27] Bala A, Malhotra S, Gupta N, Ahuja N. Emerging Green ICT: Heart Disease Prediction Model in Cloud Environment. Proceedings of International Conference on ICT for Sustainable Development, Springer, Singapore. 2016;579-87.
[28] Cloud Standards Customer Council 2017, Impact of Cloud Computing on Healthcare, Version 2.0. https://www.omg.org/cloud/deliverables/impact-of-cloud-computing-on-healthcare.htm.
[29] Markram H. The human brain project. Scientific American. 2012;306(6);50-55.
[30] Srivastav N, Urooj S. Role of Information Communication and Technology (ICT) in the Treatment of Brain Ailments. Int J Med Res Rev. 2017;5(02):119-23. doi:10.17511/ijmrr.2017.i02.04.
[31] Olesen J, Leonardi M. The burden of brain diseases in Europe. European Journal of Neurology. 2003;10(5);471-77.
[32] Cai H, Zheng VW, Chang KC-C. A Comprehensive Survey of Graph Embedding: Problems, Techniques and Applications. IEEE Transactions on Knowledge and Data Engineering. 2018; arXiv:1709.07604v3.
[33] Bruna J, Zaremba W, Szlam A, LeCun Y. Spectral networks and locally connected networks on graphs. 2013; arXiv:1312.6203.
[34] Parisot S, Ktena SI, Ferrante E, Lee M, Moreno RG, Glocker B, Rueckert D. Spectral graph convolutions for population-based disease prediction. International Conference on Medical Image Computing and Computer-Assisted Intervention, Springer, Cham. 2017;177-85.

PARTICULARS OF CONTRIBUTORS:
1. Associate Professor, Department of Electrical Engineering, College of Engineering, Princess Nourah Bint Abdulrahman University, Riyadh, Saudi Arabia. (On leave from Gautam Buddha University, Uttar Pradesh India).
2. Assistant Professor, Department of Electronics and Communication Engineering, IILM-AHL- College of Engineering and Technology, Greater Noida, Uttar Pradesh, India.
3. Software Engineer, Department of Software Development, Kalinga Institute of Information and Technology, Bhubaneshwar, Odisha, India.
4. Assistant Professor, Department of Computer Sciences, College of Computer and Information Sciences, Princess Nourah Bint Abdulrahman University, Riyadh, Saudi Arabia.

NAME, ADDRESS, E-MAIL ID OF THE CORRESPONDING AUTHOR:
Shabana Urooj,
Department of Electrical Engineering, School of Engineering, Gautam Buddha University,
Greater Noida-201312, Uttar Pradesh, India.
E-mail: shabanabilal@gmail.com; shabanaurooj@ieee.org

AUTHOR DECLARATION:
- Financial or Other Competing Interests:   Yes
- Was Ethics Committee Approval obtained for this study?    NA
- Was informed consent obtained from the subjects involved in the study?    NA
- For any images presented appropriate consent has been obtained from the subjects.   NA

PLAGIARISM CHECKING METHODS: [Jain H et al.]
- Plagiarism X-checker: Nov 30, 2019
- Manual Googling: Jan 09, 2020
- iThenticate Software: Jan 28, 2020 (12%)

ETYMOLOGY: Author Origin

Date of Submission: **Nov 29, 2019**
Date of Peer Review: **Dec 30, 2019**
Date of Acceptance: **Jan 24, 2020**
Date of Publishing: **Feb 01, 2020**